# The Fully Convolutional Transformer for Medical Image Segmentation

Athanasios Tragakis[1*] Chaitanya Kaul[2*] Roderick Murray-Smith[2] Dirk Husmeier[1]

[1]Mathematics and Statistics, University of Glasgow, United Kingdom, G12 8QW
[2]School of Computing Science, University of Glasgow, United Kingdom, G12 8RZ

## Abstract

*We propose a novel transformer, capable of segmenting medical images of varying modalities. Challenges posed by the fine-grained nature of medical image analysis mean that the adaptation of the transformer for their analysis is still at nascent stages. The overwhelming success of the UNet lay in its ability to appreciate the fine-grained nature of the segmentation task, an ability which existing transformer based models do not currently posses. To address this shortcoming, we propose The Fully Convolutional Transformer (FCT), which builds on the proven ability of Convolutional Neural Networks to learn effective image representations, and combines them with the ability of Transformers to effectively capture long-term dependencies in its inputs. The FCT is the first fully convolutional Transformer model in medical imaging literature. It processes its input in two stages, where first, it learns to extract long range semantic dependencies from the input image, and then learns to capture hierarchical global attributes from the features. FCT is compact, accurate and robust. Our results show that it outperforms all existing transformer architectures by large margins across multiple medical image segmentation datasets of varying data modalities without the need for any pretraining. FCT outperforms its immediate competitor on the ACDC dataset by 1.3%, on the Synapse dataset by 4.4%, on the Spleen dataset by 1.2% and on ISIC 2017 dataset by 1.1% on the dice metric, with up to five times fewer parameters. On the ACDC Post-2017-MICCAI-Challenge online test set, our model sets a new state-of-the-art on unseen MRI test cases outperforming large ensemble models as well as nnUNet with considerably fewer parameters. Our code, environments and models will be available via GitHub[†].*

---

[*]Equal Contribution
[†]`https://github.com/Thanos-DB/FullyConvolutionalTransformer`

## 1. Introduction

Medical image segmentation is a key tool in computer aided diagnosis. It helps detect and localise boundaries of lesions in images that can help identify potential presence of tumors and cancerous regions quickly. This has the potential to speed up diagnoses, improving the likelihood of detecting tumours and allowing clinicians to use their time more effectively, with benefits to patient outcomes [15]. Conventionally, modern medical image segmentation algorithms are built as symmetric top-down encoder-decoder structures that first compress (encode) an input image into a latent space, and then learn to decode the locations of regions of interest within images. Adding a horizontal propagation of the intermediate signal (skip connection) to this vertical information flow gives us the UNet architecture, which has arguably been the most influential leap forward in segmentation algorithms in the recent past. Most modern segmentation systems today either include the UNet or one of its variants in their pipeline. A key essence of the UNet's success is its *fully convolutional* nature. The UNet does not estimate any non-convolutional trainable parameters in its structure.

Convolutional Neural Network (CNN) based UNet models have found great success in medical image segmentation tasks in terms of accuracy and performance. However, they still require additional improvements in order to truly help clinicians in early disease diagnosis. The inherently local nature of the convolutional operator is a key issue with CNNs, as it prevents them from exploiting long range semantic dependencies from the input images. Various methods have been proposed to add global context to CNNs, most notably, the introduction of attention mechanisms, and dilating the convolution kernel in order to increase the kernel's receptive field. These methods however come with their own sets of drawbacks. Transformers have been hugely successful in language learning tasks [31] due to their ability to handle very long range sequence dependen-cies efficiently. This has led to their recent adaptation to various vision tasks [7, 18, 21, 22]. Recently proposed architectures such as ViT [7] have surpassed the performance of CNNs on benchmark imaging tasks, and many recent improvements to ViT such as CvT [36], CCT [10] and Swin Transformer [25] have shown how transformers do not need to be bulky, data hungry models, and can even work with small amounts of data to surpass the performance of CNNs. Conventionally, ViT style models first extract discrete non-overlapping patches (called tokens in NLP) from images. They then inject spatial positioning to these patches through a position encoding and pass this representation through standard transformer layers to model long rage semantic dependencies in the data.

Given the obvious merits of both CNNs and Transformers, we believe the next step forward in medical image segmentation is a fully convolutional encoder-decoder deep learning model with the ability to exploit long range semantic dependencies in medical images efficiently. Towards this goal, we propose the first *Fully Convolutional Transformer* for medical image segmentation. Our novel Fully Convolutional Transformer layer forms the main building block of our model. It contains two key components, a *Convolutional Attention* module and a fully convolutional *Wide-Focus* module (See Section 3). We formalize our contributions as follows:

- We propose the first *Fully Convolutional Transformer* for medical image segmentation, which surpasses the performance of all existing convolutional and transformer based architectures for medical image segmentation on multiple binary and semantic segmentation datasets.

- We propose a novel Fully Convolutional Transformer layer which employs a *Convolutional Attention* module to learn long range semantic context, and then creates hierarchical local-to-global context using multi-resolution dilated convolutions via the *Wide-Focus* module.

- We show, through extensive ablation studies, the effects of the various building blocks of our model in the context of their impact on model performance.

## 2. Literature review

**Early CNNs and Attention models:** The UNet [29] was the first CNN model proposed for medical image segmentation. One of the first works that introduced attention models to medical image segmentation, did it through applying a gating function to the propagation of features from the encoder to decoder of a UNet [26]. Methods such as FocusNet [17] employ a dual encoder-decoder structure where attention gating learns to propagate the relevant features from the decoder of one UNet to the encoder of the next. One of the first works that incorporates attention mechanisms inside various filter groups in grouped convolutions is FocusNet++ [19]. Many variants of UNets also exist that employ different residual blocks to enhance feature extraction [32, 28, 33, 20, 16]. UNet++ [43] creates nested hierarchical dense skip connection pathways between the encoder and decoder to reduce the semantic gap between their learnt features. Of the most influential UNet variants of recent times, the nnUNet [14] automatically adapts itself to pre-process the data and select the optimal network architecture that would be best suited to the task without the need for manual intervention.

**Transformer models:** The original Transformer architecture [31] revolutionized natural language processing tasks and has quickly become the de-facto model for visual understanding tasks as well [7]. Transformers work well for vision due to their ability to create long range visual context but suffer from the inherent drawback of not leveraging spatial context in images like CNNs. Recent works move towards possible solutions to overcome this drawback. CvT [36], CCT [10] and Swin Transformers [25] are all attempts at integrating sufficient spatial context to transformers. In medical image segmentation, most existing research looks at creating hybrid Transformer-CNN models for feature processing. Similar to the Attention UNet [26], UNet Transformer [27] enhanced CNNs with multi-head attention inside skip connections. One of the first Transformer-CNN hybrid models proposed for medical image segmentation, TransUNet [5] used a transformer encoder feeding into a cascaded convolutional decoder. Similar to TransUNet, UNETR [12] and Swin UNETR [11] use Transformers on the encoder and a convolutional decoder to construct segmentation maps. Transfuse [40] runs dual branch encoders, one with convolutional layers and the other with transformer layers and combines their features with a novel BiFusion module. The decoder for this model however, is convolutional.

**Concurrent Works** There is a recent shift from creating hybrid Transformer-CNN models, to refining the transformer block itself, to handle the nuances of medical images. Swin UNet [3] is the first architecture to propose a pure transformer to process medical images. Pure here refers to the image features being extracted and processed solely by transformer layers without the need for a pre-trained backbone architecture. DS-TransUNet [24] introduces the Transformer Interactive Fusion module to get better representations of global dependencies. Both these models have the Swin Transformer block at the heart of their computation. Concurrent works such as nnFormer [42] and D-Former [37] attempt to leverage both local and global context inside medical images through specially crafted multi-head self attention blocks to cater to this task. The main drawback with these models is their inherent linear nature of attention projection and feature processing, which FCT aims to alleviate.

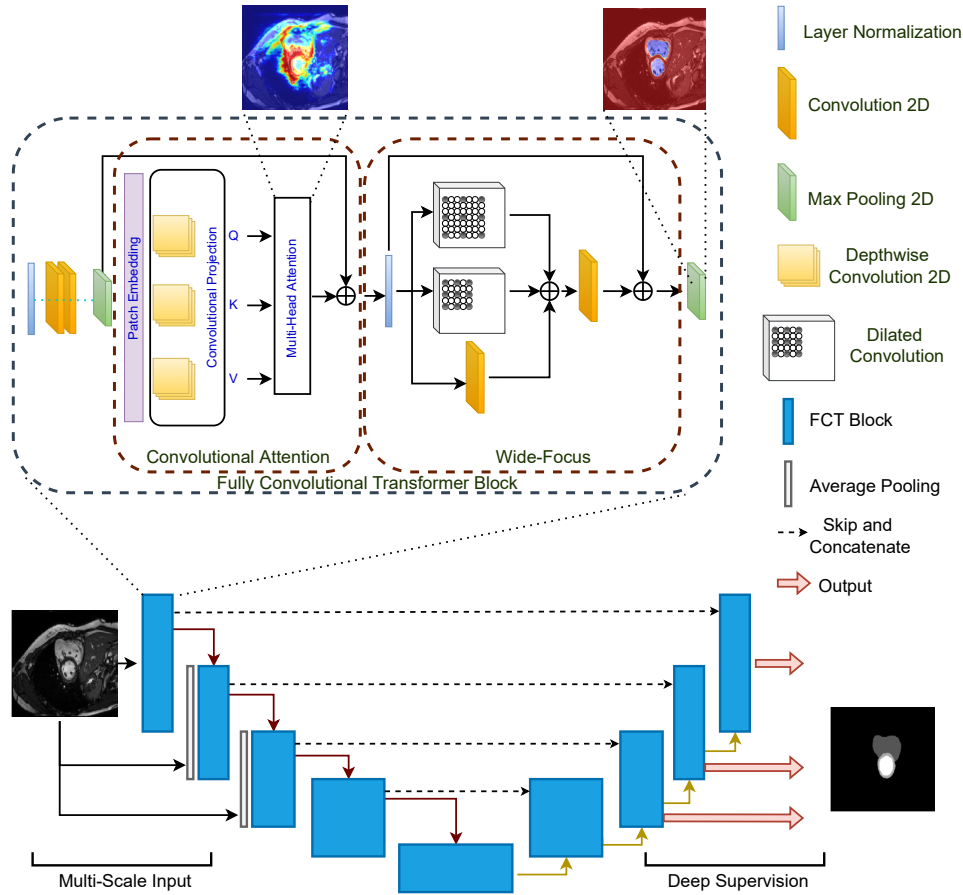

Figure 1: The *Fully Convolutional Transformer for Medical Image Segmentation*. The network (bottom) follows a standard UNet shape with the notable difference that it is purely `Convolutional-Transformer` based. The first component of the FCT layer (top) is *Convolutional Attention*. Here, `Depthwise-Convolutions` in the projection layer remove the need for positional encoding, leading to a simpler model. We create overlapping patches where the degree of patch overlap is controlled via the stride of the convolutional projection layer. To leverage spatial context from images, our MHSA block replaces linear projections with `Depthwise-Convolutions`. The *Wide-Focus* module applies dilated convolutions at linearly increasing receptive fields to the MHSA output.

Existing segmentation models in medical imaging currently suffer from at least one of three limitations. They are either based on a CNN backbone or created using convolutional layers, hence restricting their ability to look beyond their receptive fields to gain better semantic context of the images (early CNN approaches). They attempt to integrate transformers into their feature processing pipeline to leverage their ability to create long range semantic context, but in turn, make the models bulky and computationally complex (Transformer-CNN hybrids). They attempt to reduce this computational burden by creating pure transformer models for segmentation without trying to model local spatial context at a low-level feature extraction stage (concurrent works). Different from existing methods, Our *Fully Convolutional Transformer* does not suffer from these drawbacks, while still remaining a pure Transformer-based architecture for medical image segmentation. Table 4 in the supplementary material additionally summarizes the key differences of the FCT in comparison with existing works.

## 3. The Fully Convolutional Transformer

Given a dataset $\{\mathbf{X}, \mathbf{Y}\}$, where, $\mathbf{X}$ are the input images for our model, and $\mathbf{Y}$ are the corresponding semantic or binary segmentation maps. For each image $\mathbf{x}_i \in \mathbb{R}^{H \times W \times C}$, where $H$ and $W$ are the spatial resolutions of the images, and $C = \{3, ..., N\}$ are the number of input channels, our model produces an output segmentation map $\mathbf{y}_i \in \mathbb{R}^{H \times W \times K}$ where, $K \in \{1, ...D\}$. The input to the FCT is a $2D$ patch sampled from each slice of the input $3D$ image. Our model follows the familiar UNet shape, with the FCT layer as its fundamental building block. Unlike existing approaches, our model is neither a CNN-Transformer hybrid, or a Transformer-UNet structure that employs off-the-shelf transformer layers to encode or refine input features. It builds feature representations by first extracting overlapping patches from images, followed by creating a patch-based embedding of the scans and then applying multi-head self attention on those patches. The output projection of the given image is then processed via our Wide-Focus module to extract fine-grained information from the projections. Figure 1 shows an overview of our network architecture.

### 3.1. The FCT Layer

Each FCT layer begins with `LayerNormalization-Conv- Conv-MaxPool` operations. We empirically noted that applying these consecutive convolutions sequentially on the patches with a small 3×3 kernel size helps better encode image information in comparison with directly creating patch-wise projections of the image first. Each convolution layer is followed by a `Gelu` activation function. The first instance where our FCT block differs from other proposed blocks is through its application of Convolutional Attention for medical imaging.

The output of `MaxPool` is fed into a transformation function $\mathbf{T}(\cdot)$ that converts it into a new token map. Our $\mathbf{T}(\cdot)$ of choice is the `Depthwise-Convolution` operator. We choose a small kernel size of 3×3, stride s×s and a `valid` padding to ensure that, (1) the extracted patches, unlike most existing works, are overlapping, and (2) the convolution operation does not change the output size throughout. This is followed by the `LayerNormalization` operation. The obtained token map, $p_{i+1} \in \mathbb{R}^{W_t \times H_t \times C_t}$ is flattened into $W_t H_t \times C_t$, creating our patch embedded input. The next instance where our FCT layer is different from existing transformer based approaches for medical imaging applications, is through its attention projection. All existing models employ a linear position-wise projection for multi-head self attention (MHSA) computation. This results in transformer models losing spatial context, which is very important for imaging applications. Existing approaches try to alleviate this problem with convolutional enhancements to adapt them for imaging tasks. However, this adds additional computational costs to the proposed models. Inspired by the approach proposed in [36], we replace the linear projection in the MHSA block with `Depthwise-Convolutions` to reduce computational costs and leverage better spatial context from images. The patch embedding and Convolutional Attention projection form the components of our *Convolutional Attention*. Different from [36], we note that replacing `BatchNormalization` with `LayerNormalization`, helps improve performance. Furthermore, removing `Point-wise Convolutions` leads to a simpler model without losing any performance. The spatial context provided by the `Depthwise-Convolutions` further removes the need for having positional encoding, which are used to insert spatial information in the input and sequentially keep track of the position of each patch, leading to further simplifying the architecture design.

Generic transformer layers follow the MHSA block by linear layers, hence losing all spatial context in images. Directly replacing these linear layers with convolutions is a relatively straightforward approach that alleviates this problem and boosts performance. However, medical images require fine-grained information processing. Keeping this in mind, we adapt a multi-branch Convolutional layer, where one layer applies a spatial convolution to the MHSA output while the others apply dilated convolutions with increasing receptive fields to gain better spatial context. We then fuse these features via a summation and pass them through a feature aggregation layer. This feature aggregation is done through another spatial convolution operator. We call this module *Wide-Focus*. Residual connections are used to enhance feature propagation throughout the layer. The final feature is re-shaped and propagated further to the next FCT layer. Figure 1 (top) shows the FCT layer.

### 3.2. Encoder

The encoder of our model contains four FCT layers responsible for feature extraction and propagation. For the $l_{th}$ transformer layer, the output of the Convolutional Attention module is given as, $\mathbf{z}'_l = \text{MHSA}(\mathbf{z}_{l-1}) + \mathbf{z}^{q/k/v}_{l-1}$ where, $\mathbf{z}^{q/k/v}_{l-1} = \text{Flatten}(\text{DepthConv}(\text{Reshape}(\mathbf{z}_{l-1})))$. The multi-head self attention (MHSA) is denoted by, $\text{MHSA}(\mathbf{z}_{l-1}) = \text{softmax}(\frac{\text{QK}^\text{T}}{\sqrt{\text{d}}})\text{V}$. $\mathbf{z}'_l$ is then processed by the Wide-Focus (WF) module as, $\mathbf{z}_l = \mathtt{WF}(\mathbf{z}_l) + \mathbf{z}'_l$. We further inject the encoder with a pyramid style image input with the goal of highlighting different classes and smaller ROI features at different scales. It is useful to note that even without this multi-scale image pyramid input, our model is able to achieve state-of-the-art results. The (bottleneck) latent encoding of the data is created using another FCT layer.

### 3.3. Decoder

The decoder takes the bottleneck representation as its input and learns to re-sample the binary or semantic segmentation maps from this information. To create better contextual relevance in the decoder layers, skip connections from the encoder to decoder are also used where feature maps from the encoder layer at the same resolution are concatenated with the decoder layer. The decoder's shape is symmetric to the encoder. The layers in the decoder corresponding to the image pyramid layers in the encoder, output intermediate segmentation maps which provide additional supervision and boost the model's prediction ability. Contextual relevance is created by first up-sampling the feature volume and then passing it through the FCT layer to learn its best possible representation. We do not employ deep supervision at the lowest scale of FCT, and hence our model isn't 'fully deeply supervised'. This is because we observed that regions of interest (ROIs) in the input image scans were sometimes too small to segment at the lowest scale ($28 \times 28$) which resulted in a worse model performance. This low scale output added a strong bias in the model to predict some output ROIs as the background class.

### 4. Experiments

We demonstrate the effectiveness of our model through its ability to achieve state-of-the-art results across four different datasets of varying modalities. We use data from the (MRI) Automatic Cardiac Diagnosis Challenge (ACDC) [2], (CT) Synapse Multi-organ Segmentation Challenge[1], (CT) Spleen Segmentation Dataset [1] and (Dermoscopy) ISIC 2017 [6] Skin Cancer Segmentation Challenge.
The ACDC dataset contains 100 MRI scans with ground truths for the left ventricle (LV), right ventricle (RV) and myocardium (MYO). We create a `train-val- test` split of `70-10-20`. Synapse contains CT scans from 30 patients. Our experiment setup and pre-processing for Synapse is similar to TransUNet [5]. The Spleen segmentation dataset contains 41 CT volumes. Our `train-val-test` splits for this dataset are `80-10-10`. For the ISIC 2017 dataset, we create `train-val-test` splits of `70-10-20` from the 2000 images in the training dataset. We measure the performance of our models using the Dice coefficient. All the input images to our model are resized to two shapes: $224 \times 224$ and $384 \times 384$.

**Implementation Details** We run all our experiments using TensorFlow 2.0. We use one NVIDIA A6000 GPU for all our experiments. Our loss function is a equally weighted combination of the cross entropy and the dice loss. We used Adam with a learning rate of $1e-3$ which was reduced on a plateau through monitoring the validation loss. We perform warm up training for 50 epochs before training our model for a further 250 epochs. Our data augmentation is as follows: rotation (0° to 360°), zoom range (max 0.2), shear range (max 0.1), horizontal/vertical shift (max 0.3), horizontal and vertical flip. The default settings for FCT are – number of filters per stage 16, 32, 64, 128, 384, 128, 64, 32, 16, number of attention heads per stage 2, 4, 8, 12, 16, 12, 8, 4, 2. We use a batch size of 10 for ACDC, and 4 for Synapse, Spleen segmentation and ISIC 2017 segmentation. We train all our models from a randomly initialized set of weights.

### 5. Results

Our model achieves the state-of-the-art results across all reported baselines with fewer parameters and GFLOPs. FCT contains 31.7 million parameters and 7.87 GFLOPs. On the ACDC dataset, we outperform all existing works with a model size *five times smaller* than our closest competitor, nnFormer (158.92 million, 157.88 GFLOPs). We train our model on two different image sizes to see its impact on performance. As expected, FCT with a $384 \times 384$ image size achieves better results than the model with input image size $224 \times 224$, as the increased spatial resolution allows the model to see fine-grained details in the images more clearly. We also test the effect of having deep supervision across every scale vs not using deep supervision in comparison with our model. Table 2 summarizes our results on the ACDC dataset. It also shows that the deep supervision setting employed by us is the best setting for our model. To demonstrate the statistical significance of our results, we also conduct a 5-Fold Cross Validation (CV) with the ACDC dataset and compute $p$-values to show our results are significantly better than the nnFormer. We use $\text{FCT}_{224}$ for these experiments. Using 5-Fold CV, we get an average dice score of $92.43 \pm 0.38$. We then run our experiments on the ACDC dataset 5 times and average them to get a dice score of $92.88 \pm 0.09$. Both results are state-of-the-art for the dataset. Comparing with nnFormer ($91.78 \pm 0.18$), we get $p < 0.0001$

---
[1]https://www.synapse.org/#!Synapse:syn3193805/wiki/217789

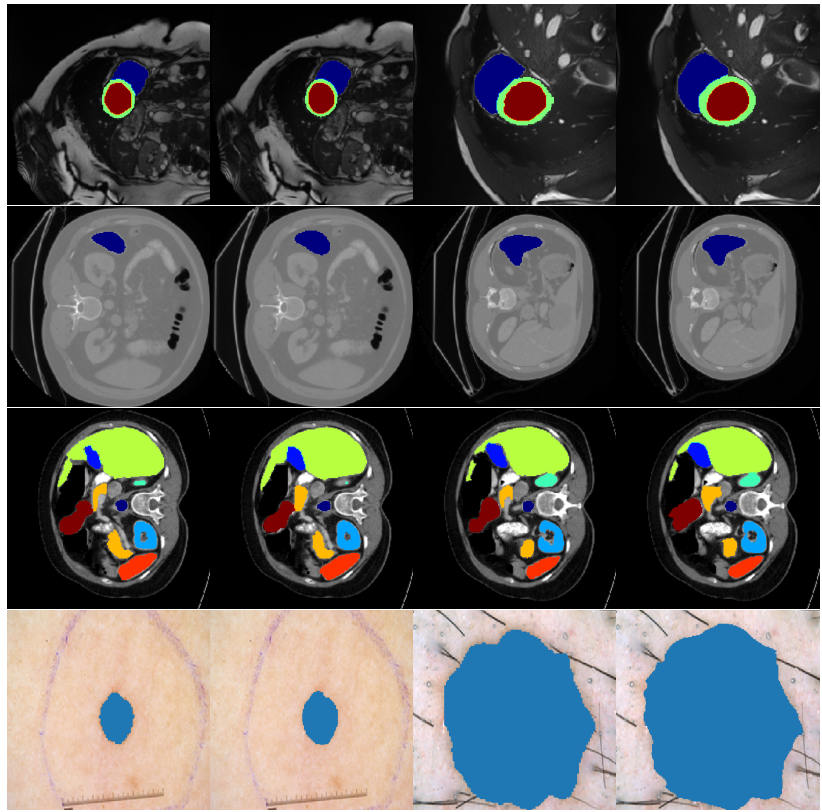

Figure 2: Qualitative results on the different segmentation datasets. From the top - ACDC Segmentation Dataset [Colours - Maroon (LV), Blue (RV), Green (MYO)], Spleen Segmentation Dataset [Colours - Blue (Spleen)], Synapse Segmentation Dataset [Colours - Blue (Aorta), Purple (Gallbladder), Navy (Left Kidney), Aquatic (Right Kidney), Green (Liver), Yellow (Pancreas), Red (Spleen), Maroon (Stomach)] and ISIC 2017 Skin Cancer Segmentation Dataset [Colours - Blue (Skin Cancer)]. The images alternate between the ground truth and the segmentation map predicted by FCT. Best viewed in colour.

in both cases, which shows the statistical significance of our results.

We compare our results on the Synapse dataset mainly with TransUNet [5], LeViT-UNet [39] and Swin UNet [3] as we use the same splits and pre-processing as those models, which suggests that any increase in performance is due to the superiority of the proposed model. We outperform all three models by considerable margins demonstrating the ability of our model to serve as a superior backbone for multi-atlas semantic segmentation tasks. TransUNet and LeVit-UNet both have ViT−12 backbones in their architecture and hence contain around 100 million parameters (and around 49 GFLOPs). Our results are summarized in Table 3.

We also achieve state-of-the-art results on the two binary segmentation tasks, Spleen segmentation (Table 1 Supplementary Material) and ISIC 2017 segmentation (Table 2 Supplementary Material). On Spleen segmentation we outperform recently proposed benchmark models such as SETR [41], CoTr [38] and TransUNet [5] by over 1.2% dice with considerably fewer parameters. On the ISIC 2017 dataset we outperform the recently proposed Boundary Aware (BA) Transformer [35] specifically designed for the task of skin cancer segmentation by 1.1% on the dice. We also evaluate the sensitivity (true positive rate) of our model, as it is a good estimate of a model's ability to accurately segment the cancer boundaries. Models trained on the ISIC 2017 dataset tend to have a high specificity but low sensitivity due to which we consider the latter here. We outperform BA Transformer on the sensitivity metric. We noticed through our ablation studies that this was largely due to the ability of our Wide-Focus module to capture hierarchical feature information at different convolution receptive fields effectively and accurately. Figure 2 shows qualitative results of our model.

**ACDC Post-2017-MICCAI Online Test Set Results.** We Train FCT (31.7 million parameters) on the 100 images in the dataset for the ACDC Challenge, and report our results on the 50 unseen test cases for which ground truth masks are not provided. We train our model on input images of size of $512 \times 512$. To account for the variations in the sizes of the images in the dataset, we crop and tile the images to get a $512 \times 512$ resolution and apply the same augmentation to the masks. To generate final predictions, we remove these extra predictions that occur due to tiling as a post processing step by averaging the tiled predictions to create the final output. We train this model as denoted in Section 4. The link to our results is available online[2] and can be compared with previous state-of-the-art results[3]. Table 4 summarises the results of the top five submissions in comparision with our results (Table 5 in supplementary material shows detailed results across all classes). We take the mean to provide average values, however, the detailed tables with full results can be found in the links provided.

| Head | Branches | Avg. | RV | MYO | LV |
| --- | --- | --- | --- | --- | --- |
| MLP | - | 91.29 | 90.5 | 88.3 | 95.1 |
| Conv1D | 1 ($D$=1) | 91.49 | 91.2 | 88.4 | 94.9 |
| Conv1D | 2 ($D$=1,2) | 91.34 | 90.5 | 88.4 | 95.1 |
| Conv1D | 3 ($D$=1,2,3) | 91.41 | 90.2 | 88.8 | 95.3 |
| Conv1D | 4 ($D$=1,2,3,4) | 91.67 | 91.1 | 88.8 | 95.1 |
| Conv2D | 1 ($D$=1) | 91.99 | 91.3 | 89.1 | **95.5** |
| Conv2D | 2 ($D$=1,2) | 91.61 | 90.9 | 88.8 | 95.1 |
| Conv2D | 3 ($D$=1,2,3) | **92.11** | **91.6** | **89.3** | **95.5** |
| Conv2D | 4 ($D$=1,2,3,4) | 91.65 | 90.6 | 89.1 | 95.2 |
| Conv2D | 2 ($k$=3,4) | 91.47 | 90.4 | 88.8 | 95.2 |

Table 1: Ablation study to determine the optimal configuration of our Wide-Focus module. $FCT_{224}$ (with 16.1 million parameters) is used for these ablations. $D$ denotes the dilation rate, and $k$ denotes the convolution kernel size.

| Method | Avg. | RV | MYO | LV |
| --- | --- | --- | --- | --- |
| R50 UNet [29] | 87.55 | 87.10 | 80.63 | 94.92 |
| R50 Att-UNet [26] | 86.75 | 87.58 | 79.20 | 93.47 |
| ViT [7] | 81.45 | 81.46 | 70.71 | 92.18 |
| R50 ViT [7] | 87.57 | 86.07 | 81.88 | 94.75 |
| TransUNet [5] | 89.71 | 88.86 | 84.53 | 95.73 |
| Swin UNet [3] | 90.00 | 88.55 | 85.62 | 95.83 |
| LeVit-UNet$_{384}$ [39] | 90.32 | 89.55 | 87.64 | 93.76 |
| nnUNet [14] | 91.61 | 90.24 | 89.24 | 95.36 |
| nnFormer [42] | 91.78 | 90.22 | 89.53 | 95.59 |
| $FCT_{224}$ w/o D.S. | 91.49 | 90.32 | 89.00 | 95.17 |
| $FCT_{224}$ full D.S. | 91.49 | 90.49 | 88.76 | 95.23 |
| $FCT_{224}$ | 92.84 | 92.02 | **90.61** | 95.89 |
| $FCT_{384}$ | **93.02** | **92.64** | 90.51 | **95.90** |

Table 2: Segmentation results on the ACDC dataset. Our model's results are reported on two different input image sizes. D.S. stands for Deep Supervision. Full D.S. is the case where D.S. is applied at every input scale.

## 6. Ablation Study

We primarily study the effect of two key components on our model's performance through ablations: removing skip connections from the encoder to the decoder, and different settings of our novel Wide-Focus module. We conduct our ablations on the ACDC dataset. Skip connections are clearly important to our model's performance (see Table 3 Supplementary Material) and the optimal setting resembles that of the original UNet [29]. To create the optimal setting of our Wide-Focus module (see Table 1), we observed the effects of wider convolutional branches and larger dilation rates on our model's performance. We observed that beyond three convolution branches with linearly increasing dilation rates, the model's accuracy starts to saturate and eventually decrease. We believe it is due to the fact that the dilated kernel fails to approximate a global kernel at the deeper layers and this leads to the dilated receptive field missing key feature information. This is also in line with our findings that smaller kernels in the FCT block lead to better performance.

---
[2]https://acdc.creatis.insa-lyon.fr/#submission/62f8e74b6a3c7704c25c679f
[3]https://acdc.creatis.insa-lyon.fr/description/results.html

| Method | Avg. | Aorta | GB | Kid. (L) | Kid. (R) | Liver | Panc. | Spl. | Stom. |
|---|---|---|---|---|---|---|---|---|---|
| R50 UNet [5] | 74.68 | 84.18 | 62.84 | 79.19 | 71.29 | 93.35 | 48.23 | 84.41 | 73.92 |
| R50 Att-Unet [5] | 75.57 | 55.92 | 63.91 | 79.20 | 72.71 | 93.56 | 49.37 | 87.19 | 74.95 |
| TransUNet [5] | 77.48 | 87.23 | 63.13 | 81.87 | 77.02 | 94.08 | 55.86 | 85.08 | 75.62 |
| TransClaw UNet [4] | 78.09 | 85.87 | 61.38 | 84.83 | 79.36 | 94.28 | 57.65 | 87.74 | 73.55 |
| LeVit-UNet$_{384}$ [39] | 78.53 | 87.33 | 62.23 | 84.61 | 80.25 | 93.11 | 59.07 | 88.86 | 72.76 |
| MT-UNet [34] | 78.59 | 87.92 | 64.99 | 81.47 | 77.29 | 93.06 | 59.46 | 87.75 | 76.81 |
| Swin UNet [3] | 79.13 | 85.47 | 66.53 | 83.28 | 79.61 | 94.29 | 56.58 | **90.66** | 76.60 |
| FCT$_{224}$ | **83.53** | **89.85** | **72.73** | **88.45** | **86.60** | **95.62** | **66.25** | 89.77 | **79.42** |

Table 3: Segmentation results on Synapse dataset. Kid. denotes Kidney, Panc. Pancreas, Spl. Spleen and Stom. Stomach. Dice Coefficient is reported.

| Method | Avg. |
|---|---|
| Mahendra Khened [23] | 91.37 |
| Georgios Simantiris [30] | 91.92 |
| Kibrom Girum [8] | 91.93 |
| Fabian Isensee [13] | 92.95 |
| Fumin Guo [9] | 93.02 |
| FCT$_{512}$ | **93.13** |

Table 4: Top 5 results on the ACDC Post-2017-MICCAI online leaderboard. FCT$_{512}$ (with 31.7 million parameters) is used for this experiment. Avg. stands for the Average dice Coefficient.

## 7. Conclusions

We proposed the *Fully Convolutional Transformer* that is capable of accurately performing binary and semantic segmentation tasks with fewer parameters than existing models. FCT is over *five times smaller* than nnFormer and *three times smaller* than TransUNet and LeViT-UNet in terms of the number of parameters. The FCT layer comprises of two key components - Convolutional Attention, and Wide-Focus. Convolutional Attention removes the need for positional encoding at the patch creation stage by using `Depthwise-Convolutions` to create overlapping patches for the model. Our `Depthwise-Convolution` based MHSA block integrates spatial information to estimate long range semantic dependencies for the first time in a medical imaging context. Wide-Focus, as seen through our ablations, helps leverage fine grained feature information present in medical images and is an important factor in boosting the performance of our transformer block. We demonstrated the ability of our model through state-of-the-art results across multiple highly competitive segmentation datasets of varying modalities and dimensions. Our FCT block is the first fully convolutional transformer block proposed for medical imaging applications, and can be easily extended to other domains and applications of medical imaging. We believe that our model can serve as an effective backbone for future segmentation tasks and pave the way for innovations in transformer-based medical image processing.

## Acknowledgements

C.K. and R.M-S. were supported by UKRI project 104690, iCAIRD, funded by Innovate UK, and from EPSRC grant EP/M01326X/1, QuantIC. R.M-S. and D.H. were also supported by EPSRC grant EP/R018634/1, *Closed-loop Data Science*. D.H. was also supported by EPSRC grant EP/T017899/1.